\documentclass{ws-ijmpa}
\usepackage{graphicx}
\usepackage{float}
\usepackage{xcolor}
\usepackage{multirow}
\usepackage{lscape}
\usepackage{tensor}
\usepackage{hyperref}
\usepackage{amsfonts}
\usepackage{amsmath}
\usepackage{amssymb}
\usepackage{graphicx}
\usepackage{dcolumn}
\usepackage[sort]{natbib}   %lo agregué para que corriera pero no era parte del texto orginal
\usepackage[normalem]{ulem}
\usepackage[papersize={8.5in,11in}]{geometry}

\DeclareMathOperator{\Tr}{Tr}

\begin{document}
%\markboth{XXXXXX et.al}{Effective textures from a $[SU(3)]^3$ flavored scalar sector}
%\catchline{}{}{}{}{}

\bibliographystyle{plainnat}%{ws-ijmpa}

%
%%%%%%%%%%%%%%%%%%%%%%%%%%%%%%%%%%%%%%%%%%%%%%%%%%%%%%%%%%%%%%%%%%%%

\title{\bf Effective textures from a $[SU(3)]^3$ flavored scalar sector}

\author{A. Carrillo-Monteverde$^\diamond$, R. Escorcia Ram\'irez$^\ddagger$, S. G\'omez-\'Avila$^\dagger$ and L. Lopez-Lozano\footnote{Corresponding author.}}
\address{\'Area Acad\'emica de Matem\'aticas y F\'isica,\\
    Universidad Aut\'onoma del Estado de Hidalgo,\\
    Carr. Pachuca-Tulancingo Km. 4.5, C.P. 42184, Pachuca, Hidalgo,
    Mexico.\\$^\diamond$alba\_carrillo@uaeh.edu.mx
    \\$^\ddagger$es339601@uaeh.edu.mx
    \\$^\dagger$selim\_gomez@uaeh.edu.mx
    \\$^*$lao\_lopez@uaeh.edu.mx}
  
\maketitle 

\begin{history}
\received{Day Month Year}
\revised{Day Month Year}
\end{history}

\begin{abstract}
Current constraints on flavor-changing neutral currents (FCNCs) strongly indicate that any new physics emerging at the 1-10 TeV scale must adhere to the Minimal Flavor Violation (MFV) principle, where Yukawa couplings are the sole sources of flavor violation. In this work, we present a model inspired by a gauged $SU(3)$ flavor symmetry that dynamically generates leptonic Yukawa matrices through effective operators. The model incorporates a scalar sector with two sets of flavons, characterized by their vacuum expectation values (VEVs), which govern the suppression scale of the Yukawa couplings and the hierarchy of neutrino masses. By leveraging phenomenologically viable Yukawa textures, we derive restrictions on the flavon VEVs and demonstrate the compatibility of the model with experimental neutrino oscillation data. Furthermore, the model predicts at least one neutrino mass to be strongly suppressed, consistent with the normal mass ordering and experimental upper bounds. This framework provides a robust mechanism for dynamically generating neutrino masses and mixing while addressing key challenges in leptonic flavor physics, such as FCNC suppression and CP-violating phases.
\end{abstract}

%\begin{abstract}
 %Current constraints on flavor-changing neutral currents (FCNCs) strongly indicate that any new physics emerging at the 1-10 TeV scale must adhere to the Minimal Flavor Violation (MFV) criterion. This means that the only sources of flavor violation in the theory should be the Yukawa couplings. Consequently, a model that explains the mass hierarchy and mixing structure of leptons must also provide a framework for the Yukawa matrices. In this work, we present one such model, inspired by a gauged-flavor framework outlined in \citep{Grinstein_2010} and further developed by \citep{Alonso_2011}. Additionally, we introduce a series of relationships between the vacuum expectation values of a set of new flavored scalars, which are essential for reproducing viable Yukawa textures \citep{Branco_2009}.

%\end{abstract}

\ccode{PACS numbers:11.30.Hv, 12.15.Ff, 13.20.-v, 14.80Cp}

%\tableofcontents
\section{Introduction}
The discovery of neutrino masses via the mixing of flavor states provides evidence of leptonic flavor violation ($L = L_e + L_\mu + L_\tau$). In the Standard Model (SM), the Yukawa sector conserves baryon number and ensures that flavor changing neutral currents (FCNC) at tree level remain diagonal for both leptons and quarks. Most of the extensions of the SM include FCNC as consequences of the introduction of new symmetries; any model beyond the Standard Model (BSM) must incorporate a mechanism to suppress FCNC at low energies in order to fulfill experimental measurements at the electroweak scale. One of the simplest approaches to achieve this is through the introduction of a flavor symmetry that guarantees the only sources of flavor violation at low energies are the Yukawa couplings \citep{SekharChivukula1987}. This concept is commonly known as the Minimal Flavor Violation (MFV) hypothesis \citep{Cirigliano_2005}.

The dynamical origin of MFV has been explored in various BSM models, ranging from technicolor to SUSY GUTs. The minimal scheme posits that the Yukawa sector arises from the spontaneous symmetry breaking of the global symmetries of the SM Lagrangian without the Yukawa sector. The global symmetry that emerges when the Yukawa couplings are canceled is referred to as horizontal flavor symmetry. In the quark sector, this symmetry can be expressed as:

\begin{equation}\label{eq:flavorgroup}
G_F \sim U(1)^3 \times SU(3)_{Q_L} \times SU(3)_{U_R} \times SU(3)_{D_R},
\end{equation}
 that has been widely studied in the literature. Generally, when low-energy flavor symmetries result from the spontaneous symmetry breaking of high-energy flavor symmetries, potentially dangerous contributions to flavor-changing processes appears \citep{Lalak2010}. These can include significant $K \overline{K}$ mixing, flavor-changing pseudo-Goldstone bosons, large electric dipole moments for electrons and neutrons, and sizable four-fermion operators generated by the exchange of light flavor quanta \citep{PhysRevD.61.116003}.

 In a work by B. Grinstein, M. Redi, and G. Villadoro \citep{Grinstein_2010}, the possibility of promoting the global flavor symmetry to a gauge symmetry is explored. In these framework, Yukawa couplings are treated as spurions that transform under the flavor symmetry, which is ultimately broken by the vacuum expectation value (VEV) of a flavon field. When this symmetry is introduced, new fermion degrees of freedom are added to cancel dangerous gauge anomalies. With this additional matter content, the Yukawa terms of the SM can emerge as a renormalizable interaction. The quantum numbers of these additional fermions are such that their mixing with the SM fermions is ``flavor-diagonal''. In this model, the simplest option to cancel non-Abelian cubic anomalies was considered. Two right-handed color fermions were added in the fundamental representation of $SU(3)_{Q_L}$, one left-handed fermion in the fundamental representation of $SU(3)_{U_R}$, and one in the fundamental representation of $SU(3)_{D_R}$. In this scheme, the masses of the new fermions are proportional to the flavon VEVs, which are in turn inversely proportional to the SM fermion masses; this inverse hierarchy provides a flavor protection mechanism such that the new physics scale can be as low as a few TeVs. Flavor protection or some stronger constraint is needed in gauged flavor theories to avoid disastrous contributions to $K-\overline{K}$ mixing and other low-energy flavor-violating observables.

The work by R. Alonso, M.B. Gavela, L. Merlo, and S. Rigolin \citep{Alonso_2011} studies the scalar potential in a general MFV model in which the global flavor symmetry $SU(3)_L \times SU(3)_U \times SU(3)_D$ is imposed, under which the u-type and d-type quarks, as well as the intergenerational doublet of $SU(2)_Y$, will transform. This model is constructed around two main ideas:
\begin{itemize}
    \item Assuming that, at low energies, Yukawa couplings are the only sources of flavor and CP violation in the SM and/or its extensions.
    \item Considering flavons transforming in the flavor symmetry that the SM exhibits in the limit of vanishing Yukawa couplings ($G_{\text{MFV}}=SU(3)_Q \times SU(3)_U \times SU(3)_D \times U(1)_{B} \times U(1)_{Y} \times U(1)_{PQ}$, where the $U(1)_{(B,Y,PQ)}$ symmetry corresponds to baryon number, hypercharge, and Peccei-Quinn symmetry).
\end{itemize}

Under the symmetry group of this theory, we have the following charge assignments: 
\begin{equation}\label{eq:fermionqn}
\begin{aligned}
Q_L &\sim (3,1,1) \\ 
U_R &\sim (1,3,1) \\ 
D_L &\sim (1,1,3). 
\end{aligned} 
\end{equation} 
Yukawa couplings are proposed as \textit{spurion fields} that transform as follows: 
\begin{equation}\label{eq:spurions}
\begin{aligned}
Y_U &\sim (3, \bar{3}, 1) \\ 
Y_D &\sim (3, 1, \bar{3}),
\end{aligned} 
\end{equation} 
preserving the invariance under the flavor group.

%Under another scheme, considering that new physics beyond the SM manifests directly only above a scale $\Lambda$, we could retain the SM as an effective theory at low energies. Since this new physics remains valid up to energies below $\Lambda$, it is not necessary for this theory to be renormalizable. Thus, it is possible to include non-renormalizable terms whose effects are suppressed by powers of $1/\Lambda^{\text{dim}-4}$.

In this work we take inspiration from both ideas to develop a flavor model for leptons addressing the origin of neutrino masses with the MFV hypothesis. First, we study a toy model that aims to capture some features of MFV in the leptonic flavor sector and write conditions to generate a hierarchical neutrino mass spectrum. A traditional approach to introduce neutrino masses introduces an effective term of dimension five ($d=5$), known as the Weinberg effective operator, given by \citep{Weinberg:1979sa}:
\begin{equation} 
\frac{\lambda_{ij}^\nu}{\Lambda_\text{NP}} (\bar{L_{Li}} H)(\tilde{H^T} L_{Lj}^C) + h.c., 
\end{equation} 
where $\lambda_{ij}^\nu$ are unknown coupling constants and $\Lambda_\text{NP}$ is the scale of New Physics (NP), where the dynamical explanation arises as a consequence of a fundamental theory beyond the SM. We have one Weinberg term for each electroweak doublet. These terms break the symmetry group:
\begin{equation} 
G_{SM}^{global} = U(1)_B \times U(1)_{L_e} \times U(1)_{L_\mu} \times U(1)_{L_\tau} \label{5u1} 
\end{equation}
which corresponds to an accidental global symmetry that arises as a consequence of the gauge symmetry and the representation of the matter fields: $U(1)_B$ corresponds to baryon number symmetry, while $U(1)_{L_e, L_\mu, L_\tau}$ corresponds to the leptonic flavor symmetries. In general, the breaking of \eqref{5u1} does not pose a problem; there is no reason why new physics should respect the \textit{accidental} symmetries of the SM. (Due to the charge assignment of the quarks it is impossible to write an equivalent term in the quark sector.)  Weinberg's term violates lepton number by 2 units and after spontaneous symmetry breaking, a bilinear term for neutrino fields is generated:
\begin{equation} 
-\mathcal{L}_{M\nu} = \frac{\lambda_{ij}^\nu}{2} \frac{v^2}{\Lambda} \bar{\nu_{Li}} \nu^c_{Lj} + h.c.,
\end{equation}
leading to a Majorana mass for left-handed neutrinos of the form:
\begin{equation} 
(M_\nu)_{ij} = \lambda^\nu_{ij} \frac{v^2}{\Lambda}. 
\end{equation}
Here we observe that the scale at which neutrino masses emerge is suppressed by $v/\Lambda$ compared to the scale of charged fermions, thus explaining their small masses. Without additional symmetries in the coefficients $(M_\nu)_{ij}$, the theory will exhibit both flavor and CP violations and the dynamical origin of the neutrino masses remains unclear.

However, we look for a model analogous to the MFV in the quark sector using an effective theory approach that does not allow for the introduction of Weinberg's effective term. This work presents exploratory research on the properties of flavored scalar sectors that can emulate the properties of Minimal Flavor Violation (MFV) theories giving a dynamical origin of leptonic masses analogous to the quark sector. We implement the symmetry 
\begin{equation}\label{eq:ours}
    G_F \sim \text{SU}_L(3) \times \text{SU}_R(3) \times \text{SU}_F(3),
\end{equation} 
and the additional degrees of freedom it entails at low energies. Here, we aim to suppress flavor violation with the MFV assumption while also proposing a mechanism to generate the Yukawa leptonic sector dynamically introducing effective terms. The dimension of the effective terms are chosen minimally to generate a hierarchical mass spectrum. 

Additional matter content will consist of flavored sets of scalar fields $\xi_L$, $\xi_R$, and $\tilde{\phi}$ that are manifested at a scale $\Lambda$, and whose transforming properties under the symmetry (\ref{eq:ours}) explain the spurion behavior (similar but not equal to (\ref{eq:spurions}) of the leptonic Yukawa matrices. Following the method of \cite{Alonso_2011} the potential is built through the Cayley-Hamilton invariants, allowing for the removal of redundant degrees of freedom that cannot be fixed by the low energy properties of the model.

Once the additional scalar fields acquire a vacuum expectation value, SSB will occur. We have made simple choices for the fermion quantum numbers that can nonetheless provide insights into the constraints required to generate viable Yukawa textures. We have made a heuristic analysis using phenomenologically viable textures to relate the vacuum expectation values of the new degrees of freedom. %Nevertheless it is possible to obtain additional restriction on the parameters when it is made an interpretation in terms of a Grand Unification Theory (GUT). In the last section we analyze some details of the relation between the MFV hypothesis and the low energy limit of a SO(10) GUT in the context of a modified  Fritzsch's texture studied in \cite{Berezhiani2024}. 

The paper is organized as follows. In section \ref{model}, we  describe the new degrees of freedom, its quantum numbers and the effective operators that generate the leptonic Yukawa sector. Also, using the Cayley-Hamilton invariants under the symmetry $G_F$ we build the potential of the flavons in order to demonstrate the existence of stable minimums. A general model for the Yukawa mass matrix is shown in terms of the flavon-like vacuum expectation values. In section \ref{texturerestrictions}, we perform a heuristic analysis of different versions of textures, specifically Fritszch-like 4 zero texture mass matrices and all those that can be obtained through a Weak Basis Transformation (WBT), in order to relate the parameters of the theory to the values of the PMNS matrix. %In section \ref{mass_spectrum} it is discussed the relation of MFV with a GUT based in a $\text{SO}(10)$ symmetry in order to reproduce some of the widely know relation of the mixing angles and ratios of the fermion masses as an application of the general scheme showed in the section \ref{model}. 
Finally, in section \ref{conclusions} we discuss our conclusions.

%****************************************************************************************
\section{Leptonic Yukawa matrices from SSB of \texorpdfstring{$G_F$}{GF}}\label{model}

The starting point is the introduction of the MFV hypothesis for the leptonic sector. We have chosen the additional flavon-like scalar fields transforming in such a way that leptonic Yukawa matrices behave as:
\begin{equation}\label{eq:spurionsleptonic}
\begin{aligned}
Y_\nu &\sim (1, 1, 8)\\
Y_\ell &\sim (1, 1, 8).
\end{aligned} 
\end{equation} 
The transformation rule of this leptonic Yukawas under the symmetry differs from equations in (\ref{eq:spurions}) because instead of the usual $G_\text{MFV}$ symmetry, we have used $G_F$, obtaining a different behavior from the quark sector. 

The minimal dynamical origin of the spurions in (\ref{eq:spurionsleptonic})  can be generated by two flavor scalar fields $\xi_{L}$, $\xi_{R}$ and a chiral scalar $\tilde\phi$. These $SU_L(2)$ singlets of the SM have the quantum numbers corresponding to the symmetry $G_\text{F}$ as shown in Table \ref{tab:Caso1}. The transformation rules for these fields are given by:
\begin{align}
    \xi'^a_{L,i}&=(U_L)^{ab} (U_{F})_{ij}\xi^b_{L,j}\label{eq:trans1},\\
    \xi'^{\tilde a}_{R,i}&=(U_R^\dagger)^{\tilde a \tilde b} (U_{F})_{,ij}\xi^{\tilde b}_{L,j}\label{eq:trans2},\\
    \tilde \phi'^{a \tilde b} &= (U_L)^{ac}\tilde \phi^{c\tilde d} (U_R^\dagger)^{\tilde d \tilde b}\label{eq:trans3},
\end{align}
with $\xi_L$ and $\xi_R$ invariant under $SU_R(3)$ and $SU_L(3)$ respectively. 

     \begin{table}[h]
       \tbl{Quantum numbers for scalars and fermions under the symmetry $G_F$.}
	{\begin{tabular}{ccccc}
 \hline
		& $\text{SU}_L(2)$ & $\text{SU}_L(3)$ & $\text{SU}_R(3)$ & $\text{SU}_F(3)$ \\ 
  \hline
		$\xi_L$ & 1 & 3 & 1 & 3 \\
		$\xi_R$ & 1 & 1 & $\bar 3$ & 3 \\
		$\tilde{\phi}$ & 1 & 3 & $\bar{3}$ & 1 \\ 
		$H$ & 2 & 1 & 1 & 1 \\
		$f_L$ & 2 & 1 & 1 & 3 \\
        $\ell_R$ & 1 & 1 &1 & $\bar{3}$\\
		$\nu_R$ & 1 & 1 & 1 & 3 \\
		\hline
	\end{tabular}}
  \label{tab:Caso1}
\end{table}

Under the charge assignments presented in Table \ref{tab:Caso1}, we construct the effective Lagrangian for the Yukawa sector up to dimension seven terms:
\begin{equation} \label{e2}
\begin{aligned}
	-\mathcal{L}_{Y_{l}} = & \bar{f_{Li}} \left[D_{ij} + \frac{(\xi_{Li}^{\dagger a} \xi_{Lj}^a+i\xi^{\dagger \Tilde{a}}_{R i} \xi^{ \Tilde{a}}_{R j})}{\Lambda^2} + \frac{\xi_{Li}^{ a} \Tilde{\phi}^{\dagger a \Tilde{b}}\xi_{Rj}^{\Tilde{b}}}{\Lambda^3}\right]f_{Rj} H,
 \end{aligned}
\end{equation}
where $H$ is the Higgs doublet and $D$ is a simple diagonal matrix to be determined by the phenomenology of the sector. In this work we will consider the case $D= \text{diag}(\alpha,\alpha,\alpha)$.

It is worth mentioning that the phase difference between the L and R flavon contribution in \eqref{e2} leads to a suppression of the cross-sections with charged leptons in addition to the usual given by negative powers of $\Lambda$. 

In this scenario flavons do not contribute to LFV; the condition $m_{\xi_R}\sim m_{\xi_L}$ allows for a smaller scale $\Lambda$ compatible with small values for Yukawa couplings in the neutrino sector. 

With the rules given in \eqref{eq:trans1}-\eqref{eq:trans3}, the Yukawa matrix transforms as
\begin{equation}\label{eq:trans4}
    Y'_{ij}=(U_F)_{ik}Y_{kl}(U_F^\dagger)_{lj}.
\end{equation}
Once the scalars fields acquire vacuum expectation values (VEVs), the effective Yukawa matrix will take the following form:
\begin{equation} \label{e4}
\begin{aligned}
        \langle Y_{ij}\rangle= & D_{ij} + \frac{1}{\Lambda^2}\xi_{Li}^{\dagger a} \xi_{Lj}^a + \frac{1}{\Lambda^2}\xi^{\dagger \Tilde{a}}_{R i} \xi^{ \Tilde{a}}_{R i} \\ 
        &+ \frac{1}{\Lambda^3}\xi_{Li}^{a} \Tilde{\phi}^{\dagger a \Tilde{b}}\xi_{Rj}^{\Tilde{b}}.
\end{aligned}
\end{equation}

The Yukawas can then be expressed as the sum of components, each transforming as \eqref{eq:trans4}, \textit{i.e.},
\begin{align}\label{eq:Yukawa}
   Y &=D+\frac{1}{\Lambda^2}(\Delta_{L}+\Delta_{R})+\frac{1}{\Lambda^3}\Gamma',
\end{align}
where the matrix elements are defined in terms of the flavon fields as

\begin{align}
    (\Delta_{L})_{ij}  &= \xi_{L i }^{a\dagger} \xi_{L j }^a\label{D12}, \\
    (\Delta_{R})_{ij}  &=  \xi_{R i }^{\tilde a\dagger} \xi_{R j}^{\tilde a}\label{D1}, \\
(\Gamma')_{ij} &= \xi^{a}_{Li} \tilde\phi^{\dagger a\Tilde{a}} \xi^{\Tilde{a}}_{Rj}\label{C9},
\end{align}
with sum over the indexes $a$ and $\tilde{a}$.

At low energies, every element of the Yukawa matrices can be expressed in terms of products of the VEVs of \(\xi_{L(R)}\) and \(\tilde\phi\). The strength of the contributions is regulated by a scale of new physics \(\Lambda\), which for simplicity we have set to the same value for the second and third terms of (\ref{eq:Yukawa}), absorbing extra couplings in the fields.
 
The effective scalar potential can be built using mass and quadratic terms for the field $\tilde\phi^{a\tilde b}$ and the Cayley-Hamilton invariants under $G_F$ for \eqref{eq:Yukawa}. With the definitions in \eqref{D12}-\eqref{C9}, we classify these invariants in terms of their mass dimension,\textit{ i.e.}, so that we obtain\citep{Alonso_2011}:
\begin{equation}\label{eq:flavorinv}
\begin{aligned}
A_{L(R)} &= \text{Tr}(\Delta_{L(R)}), \\
A_\phi &= \text{Tr}(\tilde{\phi}^\dagger \tilde{\phi}),\\
B_{\Gamma} &=\text{Tr}(\Gamma')\\
C_{LL(RR)} &= \text{Tr}(\Delta_{L(R)}^\dagger\Delta_{L(R)}),\\
C_{LR} &= \text{Tr}(\Delta_L\Delta_R).\\ 
D_{L(R)} &= \text{Det}(\Delta_{L(R)}),\\
E_\Gamma &=\text{Det}(\Gamma'),\\
\end{aligned}      
\end{equation}
where $A_i$ (for $i=L,R,\phi$) have dimension two, $B_\Gamma$ dimension 3, $C_J$ (with $J=LL,RR,LR$) dimension 4, $D_{L(R)}$ dimension 6 and $E_\Gamma$ dimension 9. Keeping the term up to dimension seven, which is the minimum required to generate three distinct nonzero masses, the flavon--only potential takes the form:
\begin{equation}\label{eq:potential}
\begin{aligned}
         V_F&= \sum_{i=L,R,\phi}\mu^2_{i} A_i+\gamma B_\Gamma+\sum_{i,j=L,R,\phi}\tilde g_{ij} A_i A_j+\sum_{J=LL,RR,LR} g_J C_J \\
        &+\sum_{i,j=L,R,\phi}\lambda_{i} A_i B_\Gamma+\sum_{i,j,k=L,R,\phi}\tilde \lambda_{ijk}A_iA_jA_k+\sum_{\substack{i=L,R,\phi\\ J=LL,RR,LR}} \lambda'_{iJ} A_i C_J\\
        &+\sum_{k=L,R}\lambda''_k D_k+\tilde\gamma B_\Gamma^2+\sum_{i,j=L,R,\phi}\sigma_{ij}A_iA_jB_\Gamma+\sum_{J=LL,RR,LR}\tilde\sigma_J C_JB_\Gamma.\\
\end{aligned}
\end{equation}
As can be seen, the presence of complex phases in the invariants leads to additional CP violation contributions. In this work, in order to make phenomenological analysis we assume that the VEVs of $\Delta_{R(L)ij}$ are real and the for $\tilde{\phi}$ being complex (see next section).
%In the following sections, we investigate the relationship between the potential and the low-energy Yukawa textures. 

%****************************************************************************************
 
\section{Restrictions from texture matrices }\label{texturerestrictions}
%\subsection{Predictions of the phenomenological viable textures}
The remarkable agreement between the Standard Model predictions and experimental results in flavor physics does not provide any insight into the nature of the ultraviolet (UV) physics that answers such open questions as the origin of flavor, the true nature of spontaneous symmetry breaking, or the family structure of quarks and leptons. Nevertheless, some phenomenological approaches to the flavor problem can be employed to parameterize UV flavor physics at low energies. One such method involves introducing \emph{ad hoc} mass matrices inspired by approximate symmetries. These matrices are characterized by the inclusion of meaningful zeros in specific entries, which respect a minimal set of flavor symmetries. These zeros, known as texture zeros, were first introduced by Fritzsch in \citep{Fritzsch:1977vd}. In this approach, a flavor basis is chosen such that the presence of zeros effectively parametrizes phenomenologically viable flavor symmetries. Such techniques have been extensively studied in various effective theories and extended models (for examples, see \citep{Grimus2004, Lam2006, Gupta2012, Ludl2014, Ludl2015}).

To illustrate how texture zeros reduce the number of free parameters and help eliminate spurious degrees of freedom, we present the relationships between the vacuum expectation values (VEVs) of the flavons introduced earlier, as classified in \citep{Branco_2009}.  The mass matrix is proportional to the Yukawa matrix $M = v \langle Y \rangle$ (where $\langle Y  \rangle$ is the VEV of $Y$). This means that the weak basis transformation (WBT) affects both matrices in the same manner, as shown in \eqref{eq:Yukawa}. To eliminate superfluous parameters from the mass matrices, it is always possible to perform a basis transformation that does not alter the physical content of the sector where the charged lepton mass matrix is diagonal, and the neutrino mass matrix takes the form of a texture. This can be expressed as follows:

\begin{equation}
    M^\ell=\text{diag}(m_e, m_\nu, m_\tau),
\end{equation}
\begin{equation}\label{eq:neutrino_masses}
    M^\nu=U^\dagger \text{diag}(m_{\nu_1},m_{\nu_2},m_{\nu_3})U=v \langle Y^\nu\rangle.
\end{equation}
where \((m_e, m_\nu, m_\tau)\) and \((m_{\nu_1}, m_{\nu_2}, m_{\nu_3})\) represent the charged lepton masses and neutrino masses, respectively, and \(v\) is the vacuum expectation value (VEV) of the Higgs boson in the case of Dirac mass terms in the Lagrangian. As can be seen from equations \eqref{eq:Yukawa} and \eqref{eq:neutrino_masses}, the matrix elements of \(M^\nu\) are suppressed by the factors \((v/\Lambda^2)\) and \((v/\Lambda^3)\). The mixing matrix \(U\) encapsulates all experimental information derived from the mixing angles defined by the Pontecorvo-Maki-Nakagawa-Sakata (PMNS) \citep{Pontecorvo:1957cp, Pontecorvo:1957qd, Maki:1962mu} parametrization, given by:
\begin{equation}\label{PMNSmatrix}
U=
\left(\begin{array}{ccc}
      c_{12}c_{13}   &  s_{12}c_{13} & s_{13}e^{-i\delta}\\
       -s_{12}c_{23}-c_{12}s_{23}s_{13}e^{i\delta}  &  c_{12}c_{23}-s_{12}s_{23}s_{13}e^{i\delta} & s_{23}c_{13}\\
       s_{12}s_{23}-c_{12}c_{23}s_{13}e^{i\delta} & -c_{12}s_{23}-s_{12}c_{23}s_{13}e^{i\delta} & c_{23}c_{13}
             \end{array} \right)P
\end{equation}
with \(c_{ij} = \cos{\theta_{ij}}\) and \(\sin{\theta_{ij}}\), where \(\theta_{ij}\) are the mixing angles and \(\delta\) is the CP-violating phase in the leptonic sector. The diagonal matrix \(P = \text{diag}(e^{i\alpha_1}, e^{i\alpha_2},1)\) contains additional phases when the Majorana terms for the neutrinos are considered, nevertheless this phases does no contributes to neutrino oscillations thus we ignore them in our analysis. It is possible to demonstrate that the element \((M^\nu)_{11}\) can be set to zero without introducing any additional assumptions or symmetries \citep{Carrillo2024}. Now, if the mass matrix is Hermitian and we change the flavor basis using the matrix:

\begin{equation}
\left(\begin{array}{ccc}
      1   &  0 & 0\\
       0&  \cos\theta & -\sin\theta e^{-i\varphi}\\
       0 &  \sin\theta e^{-i\varphi}& \cos\theta 
             \end{array} \right)
\end{equation}
The zero in \((M^\nu)_{11}\) remains unchanged, and with suitable angles, it is possible to achieve a leading non-diagonal zero texture. For instance, by introducing \((M^\nu)_{13} = 0\), one can construct a two-zero texture. These chosen angles effectively represent an assumption about the low-energy behavior of a UV flavor model. Such texture zeros can be used to impose restrictions on the VEVs of \eqref{D12}, \eqref{D1} and \eqref{C9}.

The expression in \eqref{eq:neutrino_masses} allows us to obtain some interesting restrictions using the following equations with invariants. If we define $\kappa_i$ as the eigenvalues of $v\langle Y^{\nu}\rangle$, taking into account the chiral transformations we have $2^3$ possible assignments in terms of  the masses, this is , $\kappa_i=\pm m_i^\nu$, where $m^\nu_i$ are the neutrino masses. Thus the invariant equations are given by

    \begin{align}
        \text{Tr}\left[v\langle Y^{\nu}\rangle\right]&=\kappa_1+\kappa_2+\kappa_3\label{eq:invariants1},\\
        \text{Det}\left[v\langle Y^{\nu}\rangle\right]&=\kappa_1\kappa_2\kappa_3\label{eq:invariants2},\\
        \frac{1}{2}\left[ \text{Tr}^2(v\langle Y^{\nu}\rangle)-\text{Tr}(v^2\langle Y^\nu Y^\nu\rangle)\right]&= \kappa_1\kappa_2+\kappa_1\kappa_3+\kappa_2\kappa_3\label{eq:invariants3}.
    \end{align}

\begin{figure}
    \centering
    \includegraphics[width=0.7\linewidth]{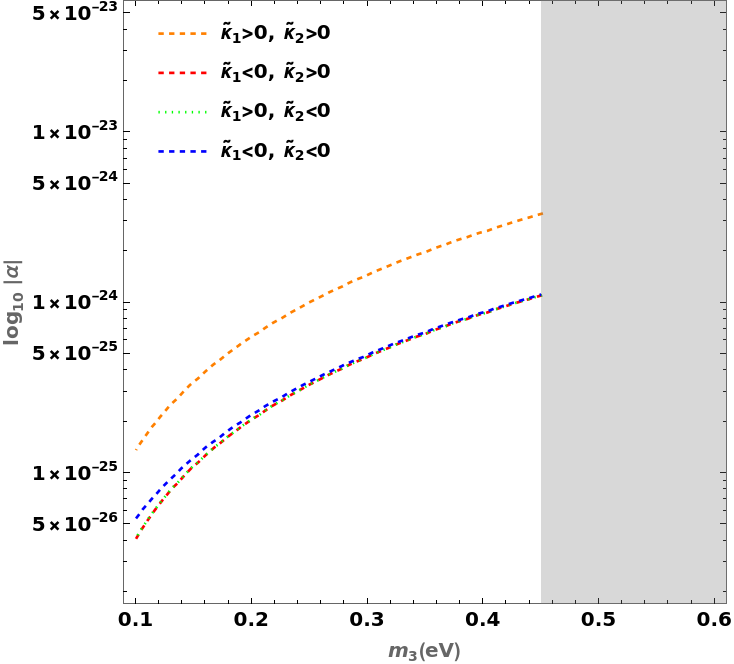}
    \caption{Behavior of the parameter scale $\alpha$ in terms of the heavier neutrino mass $m_{\nu_3}$. For small masses in $m_{\nu_3}$ the parameter $\alpha$ is highly suppressed leading to a determinant of $\tilde M_D$ very small then at least one neutrino mass is suppressed. The right-hand gray region is excluded KATRIN Collaboration \cite{Katrin2024}. Here it is defined $m_3\equiv m_{\nu_3}$}
    \label{fig:kappas}
\end{figure}

The left hand side of equations (\ref{eq:invariants1}-\ref{eq:invariants3}) can be approximated up to the dimension 7 contributions. To streamline our notation we define $\tilde\kappa_i=\kappa_i/v$, $\tilde{M}_D=\text{diag}(\tilde\kappa_1,\tilde\kappa_2,\tilde\kappa_3)$, $\Delta =\langle \Delta_L+ \Delta_R\rangle$ and $\Gamma=\langle \Gamma'\rangle$. Up to dim 7 there is a close relation between \eqref{eq:invariants1} and \eqref{eq:invariants3}, and the system of equations can be reduced to
    \begin{align}\label{}
        \text{Det}(\tilde M_D)\simeq&\alpha^3+\frac{\alpha^2}{\Lambda^2}\left(\Tr (\Delta) -\Delta_{11}\right)+\frac{\alpha}{\Lambda^3}\left(\Tr(\Gamma)-\Gamma_{11}\right)\label{bound1},\\
       \frac{1}{2}\left[ \text{Tr}^2(\tilde M_D)-\text{Tr}(\tilde M_D^2)\right]\simeq& -3\alpha^2+2\alpha \Tr(\tilde{M}_D)\label{bound2},
    \end{align}
where $\Tr(\Delta)=\Delta_{11}+\Delta_{22}+\Delta_{33}$ and $\Tr(\Gamma)=\Gamma{11}+\Gamma_{22}+\Gamma_{33}$. It is clear that up to $\mathcal{O}(\frac{1}{\Lambda^3})$ the invariants (\ref{eq:invariants1}-\ref{eq:invariants3}) restricts only the diagonal terms of $\Delta$ and $\Gamma$ matrices. %The expression \eqref{bound1} does not give and interesting restriction as \eqref{bound2}. 
We have chosen to study in this paper the Normal Ordering (NO) of neutrino masses, and also fixed $\kappa_3=m_{\nu_3}>0$; with these suppositions  the expression \eqref{bound2} defines a parametric bound for the parameter $\alpha$ that is shown in Fig. \ref{fig:kappas}. 
The behavior of the parameter $\alpha$ depends on the signs of the eigenvalues where $|\kappa_i|=m_{\nu_i}$. Using the experimental values for $\Delta m^2_{ij}$ coming from experiments of neutrino oscillation it is possible to express invariants in terms of  $\Delta m^2_{31}$,  $\Delta m^2_{21}$ and $m_{\nu_3}$. As can be seen in Fig \ref{fig:kappas} the parameter $\alpha$ is highly suppressed in the allowed interval for $m_{\nu_3}$. Taking into account the equation \eqref{bound1} this model predicts that at least one neutrino has a very small mass. The gray shaded region shows the upper experimental bound ($m_{\nu}\leq 0.45$ eV) from KATRIN collaboration\citep{Katrin2024} that use the tritium $\beta$-decay independent of any cosmological model and does not rely on assumptions on whether the neutrino is a Dirac or Majorana particle.

Additional restrictions can be found when phenomenological relations between mixing angles and masses are introduced by zero textures reflecting the misalignment of flavon vacuum expectation values. The condition to have a zero in the position $(ij)$ is given by the equation \eqref{eq:Yukawa} with $Y_{ij}=0$; therefore, texture zeros imply a relation between VEVs of the form
\begin{equation}
    \Delta_{ij}=\left(\frac{1}{\Lambda}-\frac{\Lambda^2D_{ij}}{\Gamma_{ij}}\right)\Gamma_{ij}.
\end{equation}
For non diagonal zeros, or $\Lambda^2 D_{ij}<<\Gamma_{ij}$ in the case of diagonal zeros, a linear relation $\Delta_{ij}=\Gamma_{ij}/\Lambda$ is obtained,  which reduces the free parameters of the model and suggest a particular alignment on the flavons VEVs. The elements of the Yukawa matrix can be bounded using equation \eqref{eq:neutrino_masses}, introducing the mixing matrix and masses written in terms of experimental values for $\Delta m^2_{ij}$ and $m_{\nu_3}$ as free parameter. These leads to obtain allowed regions for 
 $\Delta_{ij}$ and $\Gamma_{ij}$ through
\begin{equation} \label{PMSBound}
\left\{U^\dagger\text{diag}\left[m_{\nu_1}(\Delta m^2_{lm},m_{\nu_3}),m_{\nu_2}(\Delta m^2_{lm},m_{\nu_3}),m_{\nu_3}\right]U-v\alpha\mathbf{1}\right\}_{ij}=\frac{v}{\Lambda^2}\left(\Delta+\frac{\Gamma}{\Lambda}\right)_{ij}.
\end{equation}
As in similar models, the small masses for neutrino comes from the ratio between VEVs of flavons and the scale $\Lambda$. The presence of the parameter $v$ forces $\Lambda$ to be increased. It is possible to demonstrate that our results can be used for ample variety of texture zeros applying suitable $S_3$ transformations. The conditions given by textures matrices with two zeroes classified in terms of $S_3$ discrete transformations are described in the \ref{apendice}.

The PMNS matrix \eqref{PMNSmatrix} is in general complex, this is, contain experimental information of mixing matrices, CP violating phase for the leptonic sector, and possible Majorana phases. We proceed to determine the allowed orders of magnitude for $|\Delta_{ij}|$ and $|\Gamma_{ij}|$ using restrictions given by \eqref{PMSBound} taking $m_{\nu_3}$ on the left-hand side also as a free parameter. This allows to determine upper and lower bound for the VEVs compared with the electroweak SSB scale $v$. The mixing angles to determine the left-hand side of the equation \eqref{PMSBound} were taken from the collaboration NuFit 6.0 (2024) \cite{Esteban2020, Esteban2024} that are shown in the Table \ref{tab:parameters}. 
\begin{table}
    \tbl{Global fits taken from the collaboration NU-Fit (2006) \cite{Esteban2024} for Normal Ordering ($\Delta\chi^2=0.6$)}
    {\begin{tabular}{c|c}
    \hline
    \hline
        & $\text{bfp}\pm 1\sigma$\\
    \hline
        $\sin^2\theta_{12}$ & $0.307^{+0.012}_{-0.011}$ \\
        $\sin^2\theta_{23}$ & $0.561^{+0.012}_{-0.015}$ \\
        $\sin^2\theta_{13}$ & $0.02195^{+0.00054}_{-0.00058}$  \\
        $\delta_\text{CP}/ ^\circ$ & $177^{+19}_{-29}$\\
        $\Delta m^2_{21}$ & $\left(7.09^{+0.19}_{-0.19}\right)\times 10^{-5} \text{ eV}^2$ \\
        $\Delta m^2_{31}$ & $\left(2.534^{+0.025}_{-0.023}\right)\times 10^{-3} \text{ eV}^2$\\
    \hline
    \end{tabular}}
    \label{tab:parameters}
\end{table}
In order to find allowed regions for VEVs of flavons it is made the assumption that $\Delta_{ij}e^{\delta_{ij}}$ and $\Gamma_{ij}=|\Gamma_{ij}|e^{i\gamma_{ij}}$. The expressions that bound the VEVs are given by
\begin{equation} \label{eq:normcondition}
    \frac{v^2}{\Lambda^4}\left[|\Delta_{ij}|^2+\left(\frac{|\Gamma_{ij}|}{\Lambda}\right)^2+2\frac{|\Delta_{ij}||\Gamma_{ij}|}{\Lambda}\cos(\delta_{ij}-\gamma_{ij})\right]\simeq\left|\sum_{k=1}^3U_{ki}^*U_{kj}m_{\nu_k}(\Delta m_{31}^2,\Delta m_{21}^2,m_{\nu_3})\right|^2
\end{equation}
and for the phases
\begin{equation}\label{eq:phasecondition}
\tan\varepsilon_{ij}\simeq\frac{|\Delta_{ij}|\sin\delta_{ij}+\frac{|\Gamma_{ij}|}{\Lambda}\sin\gamma_{ij}}{|\Delta_{ij}|\cos\delta_{ij}+\frac{|\Gamma_{ij}|}{\Lambda}\cos\gamma_{ij}}
\end{equation}
where $\varepsilon_{ij}$ is a known function of mixing angles $\sin\theta_{12}$, $\sin\theta_{13}$, $\sin\theta_{23}$, the differences of squared masses $\Delta m_{31}^2$, $\Delta m_{21}^2$, the CP violating phase $\delta_\text{CP}$ and the free parameter $m_{\nu_3}$. To obtain the expressions \eqref{eq:normcondition} and \eqref{eq:phasecondition} has been dismiss the term with $v\alpha$ because, as shown in the figure \ref{fig:kappas}, is very small respect the term calculated with the PMNS matrix elements and neutrino masses. The equation \eqref{eq:phasecondition} leads to additional bounded region when it is assumed $\delta_{ij}=0$, this is the VEVs of $\Delta_{ij}$ is real and $\Gamma_{ij}$ complex, leading to additional CP violation contribution in the potential \eqref{eq:potential}.
 
\begin{figure}
    \centering
    \includegraphics[width=0.7\linewidth]{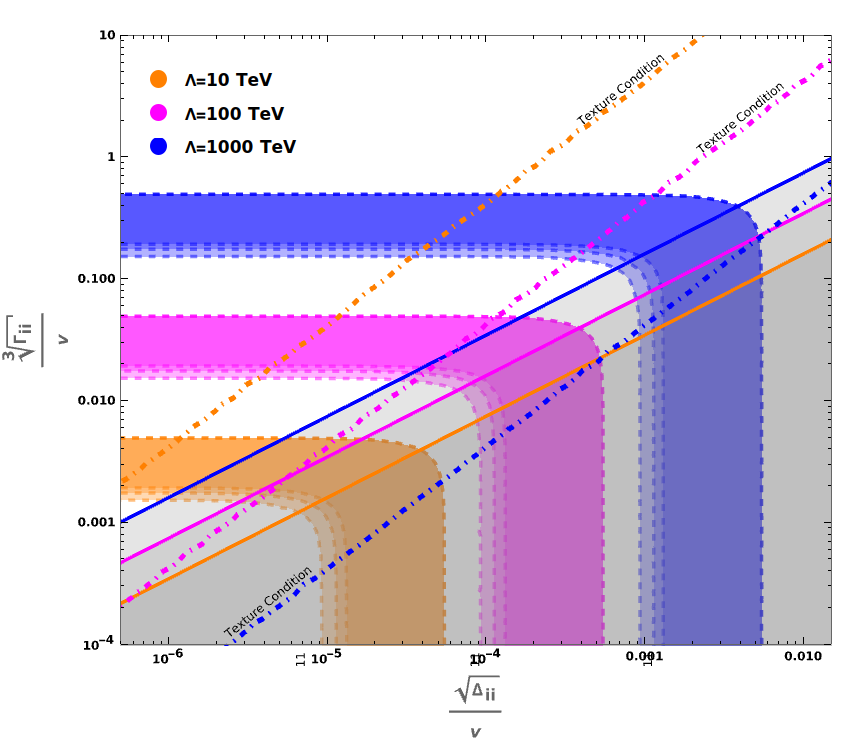}
    \caption{
    The parameter space for $\frac{\sqrt{\Delta_{ii}}}{v}$ versus $\frac{\sqrt[3]{\Gamma_{ii}}}{v}$ is shown for different values of $\Lambda$. The shaded regions indicate the allowed combinations of VEVs for $\sqrt{\Delta m^2_{31}} < m_{\nu_3} < 0.45 \, \text{eV}$. Regions below the solid color lines, which represent the constraints imposed by the $\delta_\text{CP}$ phase, are excluded. The dotted color lines correspond to restrictions imposed by a diagonal zero texture. For $\Lambda \sim 10^3 \, \text{TeV}$, the zero texture on the diagonal is excluded.
}
    \label{Diagonalflavons}
\end{figure}

\begin{figure}
    \centering
    \includegraphics[width=0.7\linewidth]{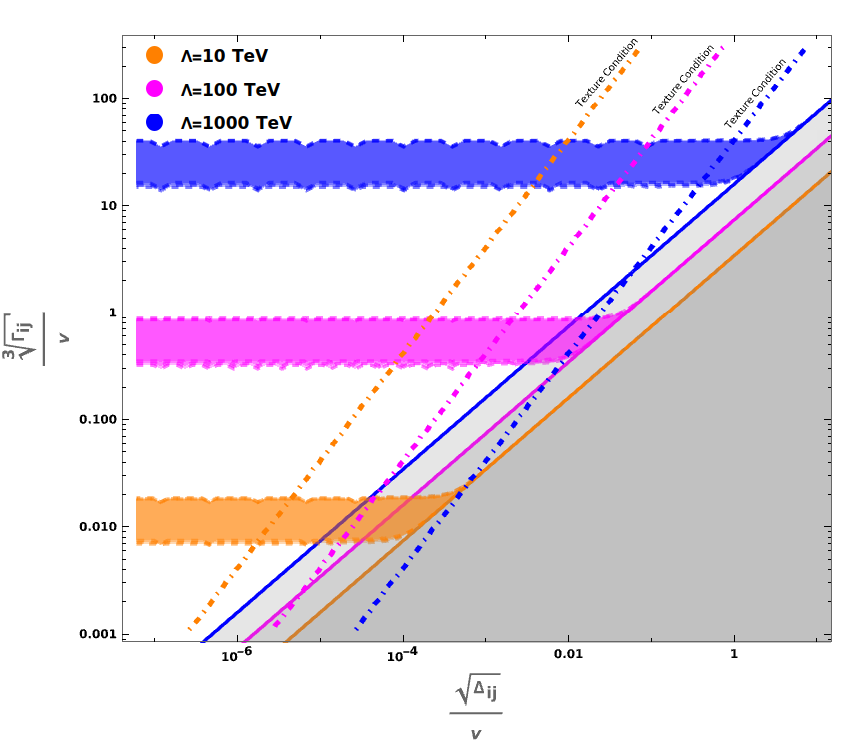}
    \caption{ The parameter space for $\frac{\sqrt{\Delta_{ij}}}{v}$ versus $\frac{\sqrt[3]{\Gamma_{ij}}}{v}$ for $i\neq j$ is shown for different values of $\Lambda$. The shaded regions indicate the allowed combinations of VEVs for $\sqrt{\Delta m^2_{31}} < m_{\nu_3} < 0.45 \, \text{eV}$. Regions below the solid color lines, which represent the constraints imposed by the $\delta_\text{CP}$ phase, are excluded. The dotted color lines correspond to restrictions imposed by a diagonal zero texture. For these no diagonal elements there non restriction on the zeros beyond those impose by the hierarchy of neutrino masses.
}
    \label{NoDiagonalflavons}
\end{figure}

In Figure \ref{Diagonalflavons} it is shown the parameter space for the diagonal VEVs for $\frac{\sqrt{|\Delta_{ii}|}}{v}$ \textit{vs} $\frac{\sqrt[3]{|\Gamma_{ii}|}}{v}$ (no sum over $i$). The color shaded regions are generated assuming normal order in the interval $\sqrt{\Delta m^2_{31}}<m_{\nu_3}<0.45\text{ eV}$ for the NP scales $\Lambda=10\text{ TeV},\ 10^2\text{ TeV}$ and $10^3$ TeV. The areas below the solid lines are excluded by the CP violation phase of the PMNS matrix, the excluded regions for different value of $\Lambda$ are shown. Also has been draw with a dotted line the restriction introduced by zero texture in the diagonal of the Yukawa matrix (see \ref{apendice} for details), in this case only the values on the dotted line are valid in order to fulfill this assumption. One of the important results that has to be pointed out is the exclusion of a diagonal zero for scales $\Lambda\sim10^3$ TeV. The blue dotted line, corresponding to the texture zero restriction is below the solid line that signal the bound coming from the $\delta_{CP}$ phase in the left-hand side of equation \eqref{PMSBound}. This exclusion results from the chosen combination of phases, \textit{i. e.} $\delta_{ij}=0$ and $\gamma_{ij}\neq 0$.
In Figure \ref{NoDiagonalflavons} it is shown the allowed regions for the no diagonal parameters. The color shaded regions are generated by the variation of the the heaviest neutrino mass in the interval $\sqrt{\Delta m^2_{31}} < m_{\nu_3} < 0.45 \, \text{eV}$ in normal order for the NP scale $\Lambda=10\text{ TeV},\ 10^2\text{ TeV}$ and $10^3$ TeV. As the previous plot it is shown with solid color lines the restriction coming from the phase $\delta_\text{CP}$ for different scales and the dotted color lines points the restriction for a zero texture. for the non diagonal elements there are no restrictions on the zero textures because all dotted color lines are above the solid lines in the allowed regions generated by the variation of $m_{\nu_3}$ 

The model studied here have many free parameters, nevertheless the phenomenology of neutrinos gives some insight on the hierarchy of VEVs of flavons. From figures \ref{Diagonalflavons} and \ref{NoDiagonalflavons} it is evident that the VEVs of $\Delta_{ij}$ and $\Gamma_{ij}$ differs by al least 2 orders of magnitude of the form $\Delta_{ij}\sim 10^{-2}\Gamma_{ij}$, thus it possible to assume that $\langle \xi^{a(\tilde a)}_{L(R),i}\rangle\sim 10^{-2}\langle\tilde\phi^{b,\hat{c}}\rangle$. 

%The restriction coming from the experimental fitting on $\Delta m^2_{23}$ and $\Delta m^2_{31}$ suggest that for our model, the flavon that carries \textit{chiral} quantum number must be light respect the one that only contain flavor quantum number as can be seen in the Table \ref{tab:parameters}. As has been studied by several authors, model that contain only fundamental representations can not generate the degeneracy of neutrino masses. In the model studied in this work the there are two sets of flavons characterized by the scale of its VEVs. The set of $\xi$'s determine the scale of the suppressed Yukawa matrices and the set of flavons $\tilde\phi$'s the hierarchy through its interactions with flavons, and the Higgs boson.

The constraints arising from the experimental fits of $\Delta m^2_{23}$ and $\Delta m^2_{31}$ suggest that, in our model, the flavon carrying a \textit{chiral} quantum number must be lighter than the flavon carrying only a flavor quantum number, as shown in Table \ref{tab:parameters}. As has been discussed by several authors, models containing only fundamental representations cannot generate the required degeneracy of neutrino masses, it is needed the bi-fundamental one. In the model analyzed in this work, there are two distinct sets of flavons, characterized by the scale of their vacuum expectation values (VEVs). The $\xi$ set determines the suppression scale of the Yukawa matrices, while the $\tilde{\phi}$ set governs the hierarchy of neutrino masses through its interactions with other flavons and the Higgs boson.

We highlight the following points:
\begin{itemize}
    \item It is assumed that the Yukawa matrix is Hermitian, implying that transitions between families in flavor processes are symmetric.
    \item The Weak Basis Transformation (WBT) allows for a reduction in the number of free parameters by introducing a diagonal zero texture, resulting in a hyper-surface in the space of VEVs of the flavon model.
    \item We chose a particular flavor basis where the charged lepton mass matrix is diagonal. Consequently, only the neutrino sector needs to be parameterized. Thus, equation \eqref{eq:neutrino_masses} can be employed to find mixing angles \(\theta_{ij}\) and \(\Delta m^2_{12}\) in terms of \((C_i, D_i, a, \Lambda)\). In principle, this can be inverted to estimate the New Physics scale under the constraints provided in Tables \ref{TipoIText} and \ref{TipoIIText-III}. Although this is not the most general case, it is enough to reproduce the PMNS matrix. 
\end{itemize}

%****************************************************************************************
\section{Conclusions}\label{conclusions}

In this study we have proposed a model for the leptonic sector mass generation based on an extended $SU(3)$ flavor symmetry, incorporating a structured scalar sector with flavons to dynamically produce viable Yukawa matrices. By leveraging the principle of MFV, this framework offers a robust mechanism to suppress unwanted FCNCs while simultaneously accommodating an explanation for the observed mass hierarchies and mixing patterns in the neutrino sector. This works presents a minimal version of the idea, where the lepton couplings are taken diagonal.

The model introduces two distinct sets of flavons, $\xi$ and $\tilde{\phi}$, whose VEVs define the suppression scale of Yukawa couplings and govern the mass hierarchy of neutrinos, respectively. Through the spontaneous breaking of the extended $SU(3)$ symmetry, effective Yukawa matrices are dynamically generated. These matrices incorporate higher-dimensional operators, reflecting the interplay between the flavon fields and the Higgs boson. Our results show that the VEVs of the $\xi$ flavons are significantly suppressed relative to those of the $\tilde{\phi}$ flavons, by at least two orders of magnitude, ensuring a natural hierarchy in the neutrino mass spectrum.

Using phenomenological constraints derived from experimental neutrino oscillation data, such as $\Delta m^2_{21}$ and $\Delta m^2_{31}$, we identified strong bounds on the parameter space of the model. Notably, the inclusion of texture zeros in the Yukawa matrices proved instrumental in reducing the number of free parameters while maintaining compatibility with experimental data. These texture zeros not only restrict the alignment of the flavon VEVs but also impose hierarchical relationships that are essential for reproducing the observed neutrino mixing angles and CP-violating phase. The results indicate that at least one neutrino mass is highly suppressed, consistent with the normal mass ordering and the experimental limits on the absolute neutrino mass scale.

We also explored the parameter space of the model in terms of the new physics scale $\Lambda$ and demonstrated that the presence of zero textures in the Yukawa matrices introduces additional constraints on the flavon VEVs. Specifically, the analysis showed that certain configurations, such as diagonal texture zeros, are excluded at high scales (e.g., $\Lambda \sim 10^3$ TeV) due to their incompatibility with the CP-violating phase constraints.

Our findings highlight the potential of the proposed SU(3)-based flavon framework to provide a consistent and predictive model for the leptonic sector. This work serves as a foundation for further exploration of flavored scalar sectors, particularly in extending the analysis to include other symmetry-breaking patterns, alternative texture configurations, or connections to cosmological phenomena like leptogenesis. Future studies could also explore the implications of this framework for charged lepton flavor violation processes or its embedding into GUTs.

In summary, this model provides a compelling approach to understanding the origin of neutrino masses and mixing, leveraging symmetry principles to address key questions in leptonic flavor physics and offering a rich avenue for continued theoretical and phenomenological investigations.

\section*{Acknowledgments}
ACM and LLL acknowledge to CONAHCYT through the SNI for partially supporting our research work and to GALC for the useful contribution and support.

\appendix\section{}\label{apendice}
The zeros in texture Yukawa matrices comes from alignment of flavons VEVs. Although there are many possible distribution of zeros, with characteristic flavor physics associated, it is possible to demonstrate that if only SM symmetries are taken into account the minimal parametrization for mass masses are given by a 2 zero texture. In this work we analyze only hermitian Yukawa matrices, thus the non-diagonal texture zeros are defined in the usual way. To generate all possible texture matrices it is defined a equivalence class with a particular texture matrix and through $S_3$ transformations generates the rest. Following the classification in \cite{Branco_2009} the 4 classes are generated with the transformation $M^{(\beta)}_i=P_i^TM_0^{(\beta)}P_i$ for $\beta=\text{I,II,III,IV}$ and $P_i\in S_3$, where
\begin{align*}
    M_0^\text{(I)}&=\begin{pmatrix}
        0&A&0\\
        A^*&e&C\\
        0&C^*&f
    \end{pmatrix};
    M_0^\text{(II)}=\begin{pmatrix}
        0&A&B\\
        A^*&e&0\\
        B^*&0&f
    \end{pmatrix};\\
     M_0^\text{(III)}&=\begin{pmatrix}
        0&A&B\\
        A^*&0&C\\
        B^*&C^*&f
    \end{pmatrix};
     M_0^\text{(IV)}=\begin{pmatrix}
        d&0&0\\
        0&e&C\\
        0&C^*&f
    \end{pmatrix}.
\end{align*}
The convention for the $P_i$ representation of $S_3$ and the restrictions coming from zero textures are given below. Here we have introduce explicitly the hermitian nature of matrix texture, thus $\{A,B,C\}$ are complex numbers and $\{d,e,f\}$ are real. The equivalence class is generated by $M^{(\beta)}_i=P_i^TM^{(\beta)}_iP_i$. 
\begin{equation*}
P_0=\begin{pmatrix}
        1&0&0\\
        0&1&0\\
        0&0&1\\
    \end{pmatrix};
    P_1=\begin{pmatrix}
        0&0&1\\
        1&0&0\\
        0&1&0
    \end{pmatrix};
    P_2=\begin{pmatrix}
        0&1&0\\
        0&0&1\\
        1&0&0\\
    \end{pmatrix}; 
\end{equation*}
\begin{equation*}
    P_3=\begin{pmatrix}
        0&1&0\\
        1&0&0\\
        0&0&1\\
        \end{pmatrix};\\
      P_4=\begin{pmatrix}
        1&0&0\\
        0&0&1\\
        0&1&0\\
        \end{pmatrix};       
     P_5=\begin{pmatrix}
        0&0&1\\
        0&1&0\\
        1&0&0\\
         \end{pmatrix};     
\end{equation*}
Thus every class contain 6 textures matrices. In Tables \ref{TipoIText} corresponding to the class I and \ref{TipoIIText-III} to the classes II and III, we present all the restrictions arising from different placements of non-diagonal texture zeros, without altering the flavor properties of the model, this is, with the same Cayley-Hamilton invariants. Also as was established in some scenarios of GUT, VEVs can be very different although it is introduced a characteristic scale $\Lambda$. the class IV is discarded because can not accommodate the phenomenological bounds. 

Although class IV is a suitable texture to explore the $\mu-\tau$ flavor symmetries \cite{Xing2014}, this analysis is restricted to those textures that allows a non degenerated mass spectra and don't produce a decoupled family, \textit{i.e.} with no more than one zero on a row (or column). For this reasons, the class IV textures are discarded. Also, it is worth mentioning that some elements of the equivalence class II and III are reiterative because some $S_3$ transformations only interchange the parameters of the texture leaving zeros at the same place. The relations coming from the PMNS matrix do not distinguish between parameters of the texture, therefore only 3 elements of those classes are included in the analysis.
\begin{table}[H]
   \tbl{Phenomenologically viable 2-zero textures matrices of class I, it is shown the alignment conditions over $\Delta_{ij}$ and $\Gamma_{ij}$. Although some relation appears two times, they are not numerically the same for every case (See text)}
   {\begin{tabular}{cccc}
     \hline
         \textbf{Texture} & \textbf{Restrictions} & \textbf{Texture} & \textbf{Restrictions} \\
         \hline
    $\begin{pmatrix}
        0 & * & 0 \\
        * & * & * \\
        0 & * & * \\
    \end{pmatrix}$ & 
$\begin{aligned}
        \Lambda \Delta_{11} &= \Gamma_{11}-\Lambda^3 \alpha,\\
        \Lambda \Delta_{13} &= \Gamma_{13}.    
\end{aligned}$
& $    \begin{pmatrix}
        * & * & *\\
        * & 0 & 0 \\
        * & 0 & * \\
    \end{pmatrix}$
& 
$\begin{aligned}
   \Lambda \Delta_{22} &= \Gamma_{22}-\Lambda^3 \alpha,\\
        \Lambda \Delta_{23} &= \Gamma_{23}.    
\end{aligned}$
    \\
%    \hline
$    \begin{pmatrix}
        0 & 0 & *\\
        0 & * & * \\
        * & * & * \\
    \end{pmatrix} $ 
& 
$\begin{aligned}
 \Lambda \Delta_{11} &= \Gamma_{11}-\Lambda^3 \alpha,\\
        \Lambda \Delta_{12} &= \Gamma_{12}.   
\end{aligned}$
&
$\begin{pmatrix}
    * & * & 0 \\
    * & * & * \\
    0 & * & 0 \\
\end{pmatrix}$
&
$\begin{aligned}
 \Lambda \Delta_{33} &= \Gamma_{33}-\Lambda^3 \alpha,\\
        \Lambda \Delta_{13} &= \Gamma_{13}.   
\end{aligned}
$    
\\
    $\begin{pmatrix}
        *  & 0 & * \\
        0 & 0 & *\\
        * & * & *\\
    \end{pmatrix} $& 
$\begin{aligned}
 \Lambda \Delta_{22} &= \Gamma_{22}-\Lambda^3 \alpha,\\
        \Lambda \Delta_{12} &= \Gamma_{12}.   
\end{aligned}$
& 
$\begin{pmatrix}
        * & * & * \\
        * & * & 0 \\
        * & 0 & 0 \\
    \end{pmatrix}$
&
$\begin{aligned}
    \Lambda \Delta_{33} &= \Gamma_{33}-\Lambda^3 \alpha,\\
        \Lambda \Delta_{23} &= \Gamma_{23}.    
 \end{aligned}$
\\
\hline
    \end{tabular}}
      \label{TipoIText}
\end{table}
%-------------------------------------------------
\begin{table}[H]
\tbl{Textures in the left-hand side corresponds to independent textures for class II, meanwhile the right-hand are of class III. The alignment condition are written for reference.}
{    \begin{tabular}{cc|cc}
    \hline
    \textbf{Texture} & \textbf{Restrictions} &\textbf{Textures} & \textbf{Restrictions}\\
    \hline
    $\begin{pmatrix}
        0 & * & * \\
        * & * & 0 \\
        * & 0 & * \\
    \end{pmatrix}$ 
    &
  $\begin{aligned}
    \Lambda \Delta_{11} &= \Gamma_{11}-\Lambda^3 \alpha,\\
    \Lambda \Delta_{23} &= \Gamma_{23}.    
 \end{aligned}$
    &
$    \begin{pmatrix}
        0 & * & * \\
        * & 0 & * \\
        * & * & * \\
    \end{pmatrix}$
    &
   $\begin{aligned}
    \Lambda \Delta_{11} &= \Gamma_{11}-\Lambda^3 \alpha,\\
    \Lambda \Delta_{22} &= \Gamma_{22}-\Lambda^3 \alpha.    
 \end{aligned}$
   \\
   $\begin{pmatrix}
        * & * & 0\\
        * & 0 & *\\
        0 & * & *\\
    \end{pmatrix}$
    &
    $\begin{aligned}
    \Lambda \Delta_{22} &= \Gamma_{22}-\Lambda^3 \alpha,\\
    \Lambda \Delta_{13} &= \Gamma_{13}.    
 \end{aligned}$
    &
    $\begin{pmatrix}
        0 &* &*\\
        * & * & *\\
        * & * & 0\\
    \end{pmatrix}$
    &
$\begin{aligned}
    \Lambda \Delta_{11} &= \Gamma_{11}-\Lambda^3 \alpha,\\
    \Lambda \Delta_{33} &= \Gamma_{33}-\Lambda^3 \alpha.    
 \end{aligned}$
    \\
    $\begin{pmatrix}
        * & 0 & * \\
        0 & * & *\\
        * & * & 0\\
    \end{pmatrix}$
     &
 $\begin{aligned}
    \Lambda \Delta_{33} &= \Gamma_{33}-\Lambda^3 \alpha,\\
    \Lambda \Delta_{13} &= \Gamma_{13}.    
 \end{aligned}$
     &
     $    \begin{pmatrix}
        * & * & *\\
        * & 0 & *\\
        * & * & 0\\
    \end{pmatrix}$
    &
   $\begin{aligned}
    \Lambda \Delta_{22} &= \Gamma_{22}-\Lambda^3 \alpha,\\
    \Lambda \Delta_{33} &= \Gamma_{33}-\Lambda^3 \alpha.    
 \end{aligned}$
     \\   
    \hline
    \end{tabular}}
    \label{TipoIIText-III}
\end{table}

\bibliography{main}
\end{document}